\documentclass[prl,superscriptaddress,twocolumn,amsmath,amssymb]{revtex4}

\usepackage{graphicx}
\usepackage{bm}

\newcommand {\Fig}[1] {Figure~\ref{#1}}

\newcommand{\beq}{\begin{equation}}
\newcommand{\eeq}{\end{equation}}

\newcommand{\pthirtyone}{$^{31}$P}

\newcommand{\beqa}{\begin{eqnarray}}
\newcommand{\eeqa}{\end{eqnarray}}

\newcommand{\ket}[1]{\left| #1 \right\rangle}

\newcommand{\tonee}{$T_{\rm{1e}}$}

\begin{document}

\title{A silicon-based cluster state quantum computer}

\author{John~J.~L.~Morton}
\email{john.morton@materials.ox.ac.uk} \affiliation{Department of Materials, Oxford University, Oxford OX1 3PH, United Kingdom}
\affiliation{Clarendon Laboratory,
Department of Physics, Oxford University, Oxford OX1 3PU, United Kingdom}

\date{\today}

\begin{abstract}
The cluster state model for quantum computation has the potential to reduce the technical challenges associated with quantum computation in certain systems. I describe how this model might be applied to donor spins in silicon with many attractive features from the point of view of practical implementation. Some of the key ingredients, such as global spin manipulation, have been robustly established, while others, such as single spin measurement, have seen much progress in recent years. A key challenge will be the demonstration of electron transfer between donors that preserves spin coherence.

\end{abstract}


\maketitle

It has been over ten years since Kane's influential proposal for a silicon-based nuclear spin quantum computer using phosphorous donors~\cite{kane98}. Since then, silicon-based architectures have been refined as the experimental challenges associated with the original proposal have become better understood~\cite{skinner03, koiller05}, while simultaneously a number of powerful and generic models for quantum computation have emerged~\cite{raussendorf05, raussendorf01}. Here, I discuss how the cluster state or ``one-way'' model for quantum computing~\cite{raussendorf01} might be advantageously applied to donors in silicon, with the potential to substantially reduce the practical requirements of a successful implementation. The essence of the scheme is to use the electron spin associated with a donor to weave an entangled network between \pthirtyone~donor nuclear spins. This resource has been shown to have exceptional coherence times~\cite{qmemory} and supports universal quantum computation through local measurements on the nuclear spins. 

The `one-way' or cluster state model for quantum computation proposed by Raussendorf and Briegel elegantly separates the act of preparing entanglement between qubits, and that of consuming entanglement in order to perform quantum computation~\cite{raussendorf01}. The orginal proposal prescribed a specific two- or three-dimensional network of entanglement, the cluster state, which is built by preparing all qubits into the $\ket{+}=(\ket{0}+\ket{1})/\sqrt2$ state, and then building graph edges through controlled-phase gates between pairs of qubits. More general topologies~\cite{universaltopology} or states~\cite{gross07} have also been shown to support universal quantum computation. Once the cluster state has been created, a circuit may be imprinted on it through selective qubit measurements in the $\sigma_z$ basis. Quantum information can then be driven through the circuit, and interactions performed, through successive $\sigma_x$ and $\sigma_y$ measurements, the results of which must all be noted. 

\begin{figure}[t] \centerline
{\includegraphics[width=3.5in]{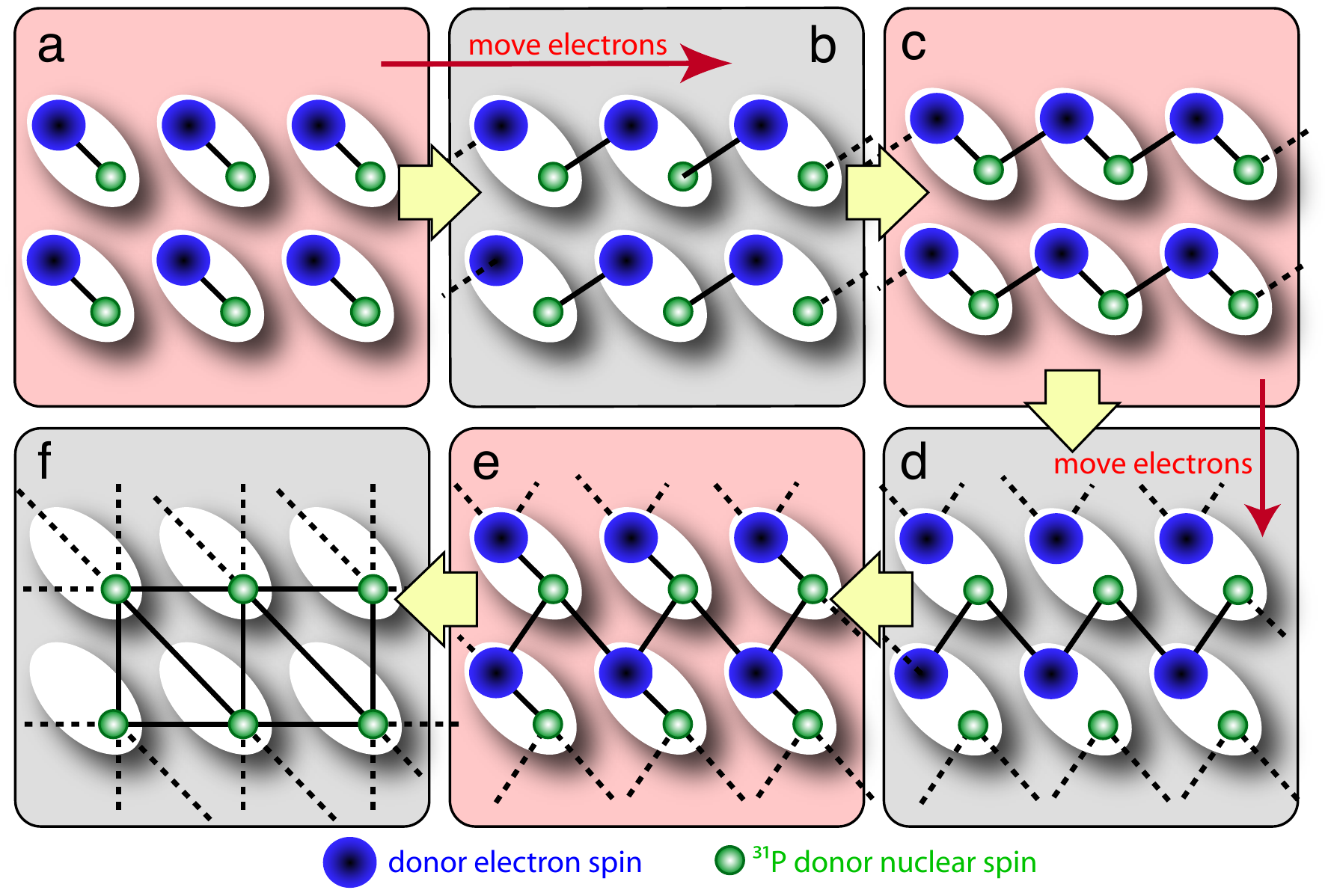}} \caption{A summary of the scheme for cluster state generation using \pthirtyone~donors in silicon. Electron and nuclear spins are offset schematically for clarity, though in reality the donor electron is localised on the nuclear spin with highly isotropic coupling. All spins are first prepared in the $\ket{+}$ state. a) A globally applied controlled-phase gate acts on all locally-coupled donor electron/nuclear spin pairs, building graph edges between them. b) Electron spins are globally moved onto the adjacent donor site. c) A second global C-phase gate is applied. d) Electron spins are moved globally onto an adjacent site, in an orthogonal direction to that of step-b. e) A final global C-phase gate is applied. f) The electron is measured in the $\sigma_y$ basis leaving the nuclear spins in a cluster state. } \label{scheme}
\end{figure}

The scheme for cluster state generation across  donor nuclear spins in silicon is summarised in \Fig{scheme}, and described in more detail below. The general requirements are i) global microwave and rf pulses, ii) the ability to globally move the donor electron to an adjacent site whilst maintaining spin coherence, iii) local measurements of each donor spin. This dispenses with the need for both gated control of spin-spin interactions and local addressability in the manipulation of spins. Recent work has shown that \pthirtyone~nuclear spins in silicon possess long coherence times (in excess of seconds)~\cite{qmemory}. The use of other donors is also possible, provided they possess a nuclear spin, such as $^{75}$As ($I=3/2$), $^{121}$Sb ($I=5/2$), $^{123}$Sb ($I=7/2$)  and $^{209}$Bi ($I=9/2$)~\cite{feher59, lo07}.

Despite their initial promise, the difficulty in preparing nuclear spins in a pure initial state has presented one of the fundamental barriers to scaling up nuclear spin quantum computing implementations~\cite{vandersypen01, Warren1997}. Electron spins benefit from a much larger magnetic moment than nuclear spins and can be cooled into a ground state of high purity at experimentally accessible fields and temperatures. Electron spins can thus provide a resource to cool the nuclear spin, through a SWAP operation between electron and nuclear spin which has been demonstrated in the context of a quantum memory~\cite{qmemory}. Waiting several \tonee~or reloading the electron spin then re-purifies the electron spin and both spins have been cooled into a ground state. Alternative schemes based on dynamic nuclear polarisation through optical excitation have already shown electron and nuclear spin polarisation as high as 90$\%$ and 76$\%$ respectively~\cite{mccamey09,yang09}. In order to prepare the spins in the desired $\ket{+}$ state, global resonant microwave (rf) $\pi/2$ pulses are applied to the electron (nuclear) spins. 

\begin{figure}[t] \centerline
{\includegraphics[width=3.5in]{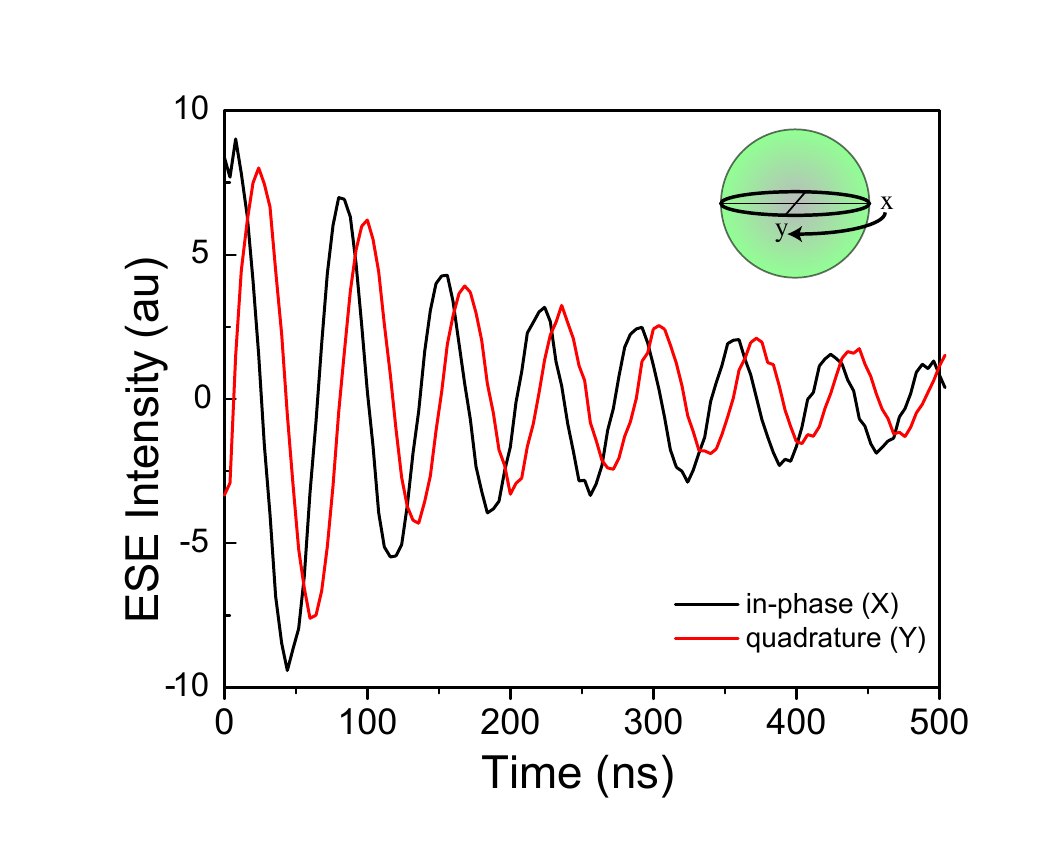}} \caption{Experimental demonstration of controlled Z-gate between an electron and a nuclear spin of a \pthirtyone~donor in Si, using an ensemble of $\approx 10^{13}$ spins. This operation with $\theta=\pi$ forms the entangling operation that builds the cluster state, and takes $\approx40$~ns in this experiment. Temperature: 8~K.} \label{zgate}
\end{figure}

The key entangling operation exploited here is the controlled-phase gate between the electron and nuclear spin located on the same donor site, coupled via an isotropic hyperfine coupling  ($\approx 120$~MHz)~\cite{feher59}. This can be performed through a selective Z-rotation of either spin, depending on the state of the other. Electron spin manipulation can be performed on a much faster timescale than that for nuclear spin, so it is advantageous to implement this gate through a $\pi$ Z-rotation of the electron spin, selective on the state of the nuclear spin. As the hyperfine coupling is well-resolved in this system, such a selective rotation can be simply applied using, for example, the composite rotation $\left(\frac{\pi}{2}\right)_x  \left(\pi\right)_y  \left(\frac{\pi}{2}\right)_{-x}$. \Fig{zgate} shows the implementation of such a controlled phase gate using the electron and nuclear spins of Si:\pthirtyone, with a gate operation time of 40~ns. The gate is simultaneously performed on $\approx 10^{13}$ spins, illustrating that global control with high fidelity is possible.

Although the entangling gate described above can be applied with high fidelity using established techniques, it does not of itself offer a scalable route to generating an entangled cluster state network. Using direct spin-spin (e.g. dipolar or exchange) interactions between spins in adjacent donors is fraught with difficulty as it relies on atomic-scale precision in the positioning of donors within the silicon substrate. Such interactions would be always-on, and the addition of further top gates to control the interaction, as proposed in the original Kane model~\cite{kane98}, brings yet more experimental challenges such as gate alignment. However, the fact that the donor can be ionised and the electron spin moved onto an adjacent donor site can be exploited~\cite{andresen07}. This formed the basis for the two-qubit gate proposed in Ref~\cite{skinner03}. This process must preserve the coherence of both the electron and nuclear spins, such that the pair remain entangled although they have become spatially separated. Spin-coherent transport in silicon over 350~$\mu$m has been observed through a wafer sandwiched between ferromagnetic thin films for spin injection and detection~\cite{huang07,appelbaum07} and control of electron shuttling between a pair of donors 50~nm has been reported~\cite{andresen07}. 


There remain many open questions as to the best way to implement the key shuttling step, and what measures should be taken to ensure spin-coherence. However, the global nature of this operation (i.e. all donor electrons shift along $x$, or $y$) simplifies the practical implementation considerably and permits the generation of the complete cluster state resource in just two shifting steps. 

Coherent charge transfer across donors has been proposed using an intermediate donor and gated control of inter-donor coupling~\cite{greentree04}. This could be used to move electrons in the scheme proposed here, however maintaining charge coherence is not necessary here. The application of global potentials coupled with some optical excitation to promote the donor into the conduction band, provides another route to moving electrons between donor sites. The high energy of the double occupancy state of the donor may lead to some cooperativity in the transfer process along a linear change of donors. 

The use of a 3-phase CCD architecture, would provide greater control in inter-donor shuttling. As we do not exploit, or even desire, direct interactions between donor spins, separations in excess of 500~nm are preferable and provide plenty of room for the necessary control gates. Shuttling times will vary depending on the method used, but shuttling rate of 1~MHz provides a conservative estimate. The three controlled-phase gates have a total time of order 100~ns, while the growth time of the $N$-qubit cluster state scales only with $\sqrt{N}$, giving a preparation time of order 100$\mu$s for a 10$^4$ qubit device. Alternatively, if donors were prepared such that alternate ones were ionised, the shuttling can be simultaneous rather than sequential, via the intermediate (and empty) donor sites. Thus the growth time of the cluster state would be largely independent of $N$ (neglecting the effect of errors which is discussed in a following section), and could be on the order of a few microseconds.

It may be advantageous to apply dynamic decoupling of the electron and nuclear spin throughout the transfer process, so as to make the nuclear spin immune to uncertainties in the transfer time, however the large hyperfine coupling strengths found in many donors would make this challenging.

The nature of measurement-driven quantum computation is such that the quantum logic gate operation speed and fidelity are directly determined by the quality of qubit measurement. This scheme demands local measurement of each donor nuclear spin which is fast and accurate.  Spin-dependent transport has been proposed as a method for nuclear spin measurement~\cite{sarovar07}. The spectral position of the (electrically detected) electron spin resonance forms a measurement of the nuclear spin state~\cite{boehme,mccamey06, lo07, beveren08}. Alternatively, the nuclear spin state may be SWAPped into the electron spin, which can then the measured using a variety of techniques proposed, including spin-dependent tunneling into a single electron transistor~\cite{hansapl, morello09}.

Measurement times have so far been long (seconds), and it will be critical to reduce these. Nevertheless, spin measurement techniques in silicon are undergoing highly active development and it is reasonable to expect these times to fall. Using experimentally determined values of nuclear coherence time, spin measurements at $\approx$40~kHz would yield a figure of merit of $10^5$ (coherence time/`gate' time).

Strictly, we require the ability to perform measurements in the $\sigma_x$, $\sigma_y$ or $\sigma_z$ basis. This could be achieved through a combination of $\sigma_z$ measurements with single qubit manipulation prior to measurement. However, given global spin qubit manipulation, all spins would be manipulated far more often than if local addressability was permitted. Given the high-fidelity with which nuclear spins may be manipulated, this is not immediately problematic, but it will provide a source of error which scales with the size of the cluster state. It might be advantageous to investigate methods for measuring spins in a locally chosen basis, or use a magnetic field gradient for spatial addressability such as is used in magnetic resonance imaging, or ion traps chips~\cite{wang09}.

The first measurement to be performed is to project out the electron spin to leave the donor nuclear spins in the cluster state. This is important to make the most of the long nuclear spin coherence time. A measurement of the electron spin in the $\sigma_y$ basis builds graph edges between all three of nuclear spins with which it was entangled, as long as the measurement outcome for each spin in known. The resulting graph state of nuclear spins can be projected into a topology capable of universal quantum computation, for example a hexagonal lattice~\cite{universaltopology}. The precise ordering of entangling, shuttling and measuring operations can be modified to yield other cluster state structures. For example, by performing an additional $\sigma_y$ electron spin measurement and C-phase gate before the second shuttling step (between c and d in \Fig{scheme}), a square graph state lattice can be constructed.

There are number of different manifestations of error in this architecture. The first will be fabrication errors where there is a `dead pixel' such as a missing donor, faulty spin measurement device, or nearby charge trap. These can be diagnosed in advance of any calculation and the pixel can be avoided when projecting the cluster state into a graph state for a particular quantum algorithm. A second kind of error arises in the construction of the cluster state, or measurement errors. Finally, there is decoherence of the nuclear spin qubits which comprise the cluster state. The coherence time of \pthirtyone~nuclear spins in silicon have been measured in excess of seconds at 5.5~K. The removal of the electron spin from the donor nuclei is likely to extend this time further and to higher temperatures, however, elements such as the spin-detection device may require sub-1K temperatures, regardless. 

The mitigation of all such errors remains the subject of much study, however, in all cases robust solutions to error-correction have been proposed, see for example, treatments of hole defects~\cite{kieling07}, and schemes for qubit loss which tolerate errors up to 50$\%$~\cite{varnava06} . Error correction in cluster states have been examined both in terms of adapting conventional fault-tolerance approaches~\cite{steane03,nielsen05} and in designing novel approaches specific to the cluster state model using topological codes~\cite{goyal07}. The topological fault tolerance scheme has a high error tolerance, which comes at the cost of a large overhead. However, the inherent scalability of this silicon-based architecture makes it ideal for the implementation of such codes.

The different elements of the silicon-based cluster state scheme described here are at different stages of maturity. Global spin manipulations, both at the single and two-qubit level have been well studied. Single spin measurement of donors in silicon is of wide interest and given current progress it is likely that this will be solved in the very near future. The element which deserves the most attention will be the spin-coherent shuttling of donor electron electron spins between adjacent donor sites, which is fundamental to the construction of the cluster state resource.

This work has benefitted substantially from stimulating  discussions with Stephen Lyon, Simon Benjamin, Dane McCamey, Hans Huebl and Joe Fitzsimons. I acknowledge support from St. John's College, Oxford and from the Royal Society. 

\bibliography{clustersi}

\end{document}